%% file: rev2paper.tex
\documentclass{aa}
\usepackage[dvips]{graphicx}
\usepackage{psfig}

\newcommand{\lapp}{\mbox{\raisebox{-0.3em}{$\stackrel{\textstyle <}{\sim}$}}}

\begin{document}

\thesaurus{}

\title{Radio observations of the circumnuclear ring in NGC6951}

\author{ D.J. Saikia \inst{1,2,3} \and Bikram Phookun \inst{4} \and
A. Pedlar \inst{2,5} \and  K. Kohno \inst{6,7}  }

\titlerunning{The circumnuclear ring in NGC6951}

\institute{
Department of Physics, Queen's University, Kingston, Ontario, K7L 3N6, Canada 
\and
Jodrell Bank Observatory, University of Manchester, Macclesfield, Cheshire SK11 9DL, UK
\and
National Centre for Radio Astrophysics, 
Tata Institute of Fundamental Research, Post Bag No. 3,\\
Ganeshkhind, Pune 411 007, India 
\and
St. Stephen's College, University of Delhi, Delhi 110 007, India 
\and
Onsala Space Observatory, S-43992 Onsala, Sweden
\and
Nobeyama Radio Observatory, Minamimaki, Minamisaku, Nagano, 384-1305, Japan 
\and
Institute of Astronomy, University of Tokyo, Osawa, Mitaka, 181-0015, Tokyo, Japan
}

\date{Received 26 September 2001; Accepted 3 December 2001}

\maketitle

\begin{abstract}
We present sensitive, high-resolution radio observations of the circumnuclear region of 
the barred spiral galaxy NGC6951. These observations reveal a ring of radio emission 
with many discrete components and a marginally resolved nuclear component. We compare
the radio ring with observations at other wavelengths, and discuss the nature of the
compact radio components. 
\end{abstract}

\keywords{galaxies: active  -- galaxies: individual: NGC6951 -- galaxies: nuclei -- galaxies: spiral -- 
radio continuum: galaxies }

\section{Introduction }
The dynamics and relationships between gas inflow processes,
circumnuclear star formation and the presence of an active nucleus are important,
since they might provide valuable leads towards understanding the transport of
gaseous material towards the nuclear regions of active galaxies. An interesting class of
objects for studying these aspects are the S\'{e}rsic-Pastoriza or S-P galaxies
(S\'{e}rsic \& Pastoriza 1965; S\'{e}rsic 1973) which include galaxies with diffuse 
and amorphous nuclei in addition to the well-known hot-spot systems. 
S\'{e}rsic \& Pastoriza (1967) found that galaxies with these unusual nuclear morphologies 
occurred in barred or mixed-type galaxies. The S-P 
galaxies reflect a broad spectrum of properties with some of them also
harbouring an active nucleus. The existence of both starburst and an active nucleus
makes them particularly interesting for studying their kinematics, 
evolution and possible relationships between the two forms of activity. 
Circumnuclear star formation has been seen in many hot-spot galaxies, and
there have been suggestions of a correlation between circumnuclear rings and nuclear
activity (cf. Arsenault 1989). 

Theoretical studies suggest that circumnuclear rings arise due to a bar-driven
inflow of gas and dust to an inner Lindblad resonance (ILR) or between two ILRs
(e.g. Combes \& Gerin 1985; Athanassoula 1992; Byrd et al. 1994; Piner et al. 1995).
The dense gas which accumulates in an ILR ring leads to a high star-formation
rate either due to collisions of the molecular clouds (Combes \& Gerin 1985), or
gravitational collapse in the ring when the density reaches a critical value
(Elmegreen 1994). To understand the formation and fuelling of the AGN, the gas
must flow inwards from the ILR to sub-parsec scales, and this is at present
not well understood (cf. Axon \& Robinson 1996). Some simulations suggest small, 
steady inflow (Piner et al. 
1995), while others suggest rapid inflow for brief periods (Wada \& Habe 1992;
Heller \& Shlosman 1994).  

In this paper we report high-resolution radio observations of the nuclear
region in the barred, late-type grand-design galaxy NGC6951, which we had earlier
observed as part of a survey of 47 S-P galaxies with the Very Large Array (VLA) 
at $\lambda$20 and 6 cm
(Saikia et al. 1994). The observations reported here are of higher resolution and
better sensitivity, and clarify the radio structure of the circumnuclear region 
of this well-studied galaxy. New supernovae have been reported in this galaxy
during 1999 and 2000 (Cao et al. 1999; Valentini et al. 2000). 

\begin{figure*}[t]
\vspace{0.2in}
\hbox{
\hbox{
\psfig{figure=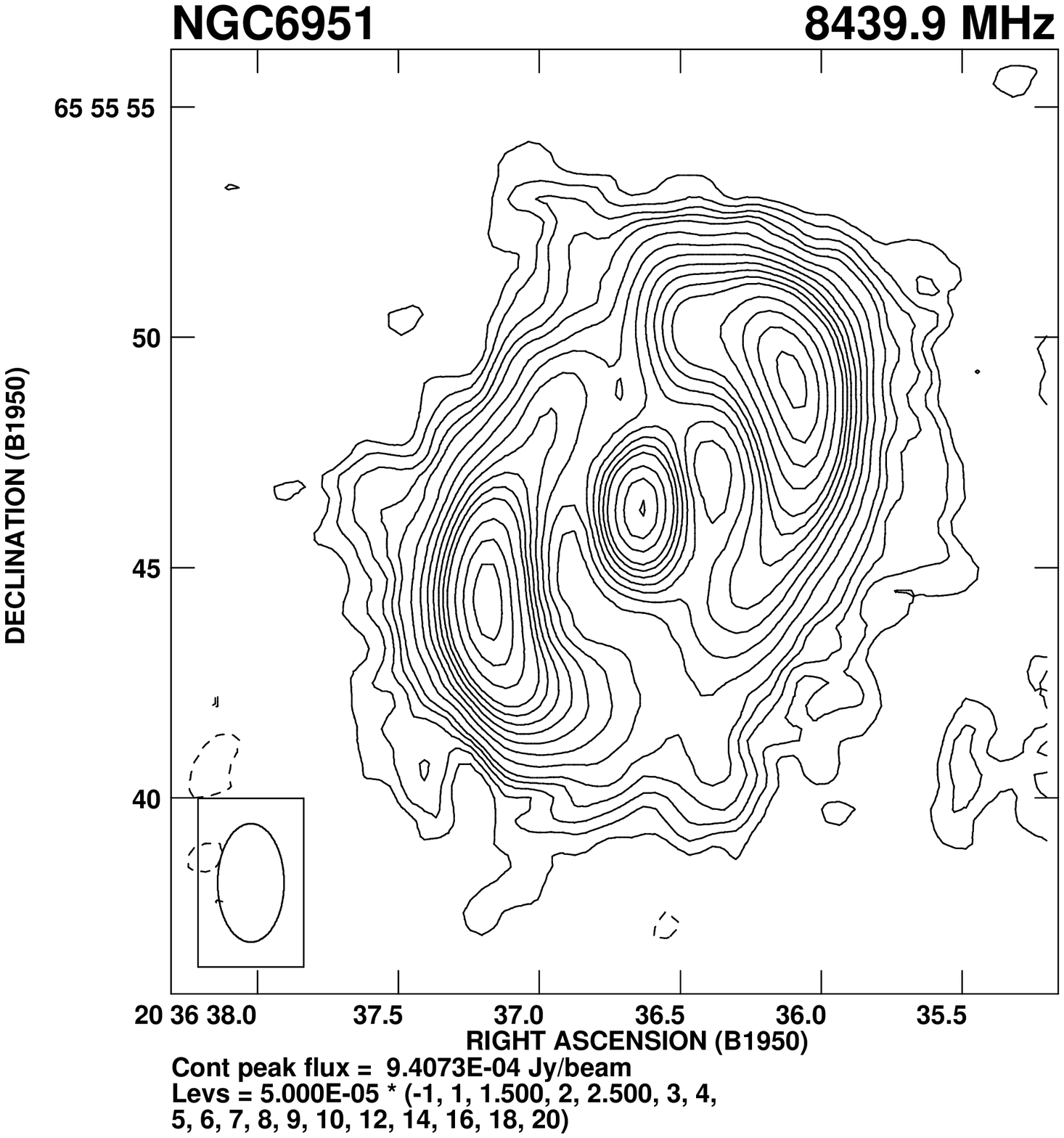,width=3in}
}
\hspace{0.8in}
\hbox{
\hspace{-0.7in}
\psfig{figure=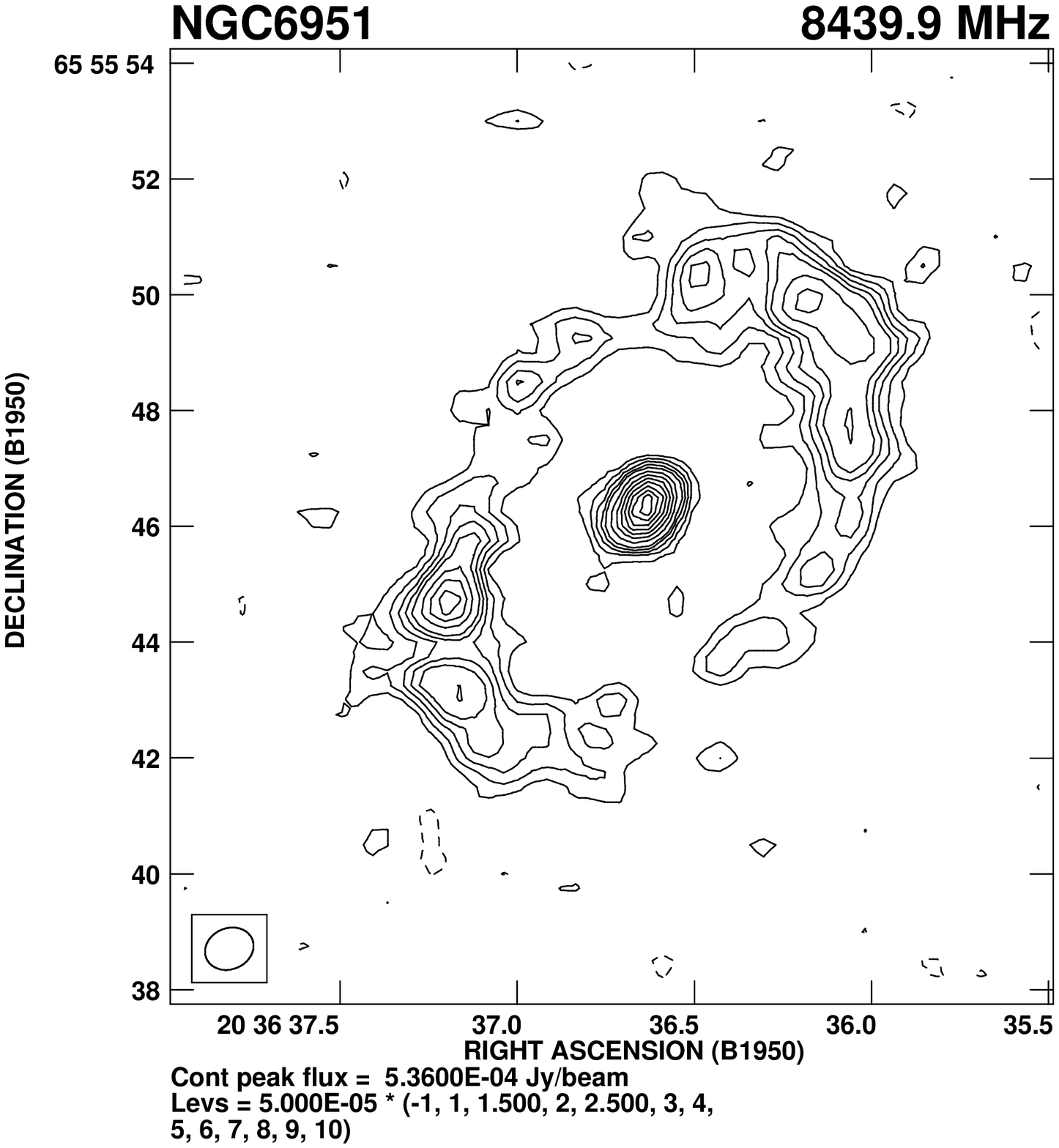,width=3in}
}
}
\caption{Radio images  of the nuclear region of NGC6951 at  8.4 GHz with an 
angular resolution of 2.57 $\times$ 1.44 arcsec$^2$ along PA 0$^\circ$ (left panel), and 
0.86 $\times$ 0.71 arcsec$^2$ along PA 113$^\circ$ (right panel). The peak 
brightness and contour levels are indicated below each image. 
}
\end{figure*}
 
\section{NGC6951}

NGC6951, which has been classified as a spiral galaxy of type SAB(rs)bc (de Vaucouleurs et al.
1991), has an active nucleus (Boer \& Schulz 1993; Ho et al. 1995) and a circumnuclar 
ring of star formation (Buta \& Crocker 1993; Barth et al. 1995). The active nucleus
has been classified either as a Seyfert 2 (e.g. Boer \& Schulz 1993; Ho et al. 1995, 
1997a) or a LINER (Filippenko \& Sargent 1985; Mu\~{n}oz-Tu\~{n}\'{o}n et al. 1989; M\'{a}rquez \& Moles 
1993). P\'{e}rez et al. (2000) have considered the available data and have classified it as
a transition object between a `very high excitation LINER  and a possible nitrogen 
over-abundant Seyfert 2.' 

It is at a distance, d, of 24.1 Mpc (Tully 1988), so that 1$^{\prime\prime}$ corresponds to 
117 pc. M\'{a}rquez \& Moles (1993) have noted its high degree of isolation with no other 
galaxy within a projected distance of 1 Mpc and 500 km s$^{-1}$ in redshift, suggesting that it has 
been free of any gravitational influence for about 10$^9$ yr. 

There have been a large number of studies of this galaxy which include broad- and
narrow-band observations at optical and infrared wavelengths (Buta \& Crocker 1993;
M\'{a}rquez \& Moles 1993; Barth et al. 1995; Wozniak et al. 1995; Elmegreen et al. 1996;
Friedli et al. 1996; Knapen 1996; Rozas et al. 1996a,b; Gonz\'{a}lez Delgado
et al. 1997; Gonz\'{a}lez Delgado \& P\'{e}rez 1997; Mulchaey et al. 1997; Elmegreen et al. 1999;
P\'{e}rez et al. 2000), spectroscopy of the nuclear region (Filippenko \& Sargent 1985;
Mu\~{n}oz-Tu\~{n}\'{o}n et al. 1989; Boer \& Schulz 1993; M\'{a}rquez 
\& Moles 1993; Ho et al. 1995, 1997a,b;
P\'{e}rez et al. 2000), radio images of the nuclear region (Vila et al. 1990; Saikia
et al. 1994; Ho \& Ulvestad 2001, hereinafter referred to as HU) and high-resolution 
images of the molecular gas (Kenney et al. 1992; Kohno et al. 1999). 

The most prominent features are the bar, the bulge and the spiral arms, with the bar 
being better defined at infrared wavelengths. The circumnuclear region has a pseudo
outer, pseudo inner and circumnuclear ring, with evidence of isophotal twisting 
between the bar and the circumnuclear ring. The dust lanes appear as straight features
from the bar edges. The ring morphology has many bright knots, with the largest 
concentration being towards the north-western and south-eastern sides of the ring. 
The hotspots or knots of emission are also seen in the infrared K-band and there is a 
general correspondence between the positions of the hotspots in H$\alpha$ and in 
K band, although in some cases there is some offset in the positions (Knapen 1996). 
The CO (1$-$0) emission shows a twin-peak morphology which could be due to orbit crowding near 
the outer ILR (Kenney et al. 1992; Kohno et al. 1999). The HCN observations by Kohno et al. show
this dense gas to also have twin-peaks, but these are offset from the CO peaks and appear 
to be closely related to the radio peaks in the lower-resolution images.  

\input{table1.tex}

\section{The nuclear radio ring in NGC6951}

In Fig. 1 we present the convolved and full-resolution radio images
of NGC6951 at 8.4 GHz. These observations were made with the VLA B-array
on 1995 October 30 and have an angular resolution of 2.57 $\times$ 1.44 arcsec$^2$
along position angle (PA) 0$^\circ$, and 0.86 $\times$ 0.71 arcsec$^2$ along PA 113$^\circ$
respectively. The rms noise level in each of the images is $\sim$0.02 mJy/beam.

The low-resolution image shows two prominent outer peaks at RA 20$^h$ 36$^m$ 
36.$^s$10, Declination 65$^\circ$ 55$^\prime$ 49.$^{\prime\prime}$00, and 
RA 20$^h$ 36$^m$ 37.$^s$20, Declination 65$^\circ$ 55$^\prime$ 44.$^{\prime\prime}$25,
with peak flux densities of 0.94  and 0.86 mJy/beam respectively. The peak flux density
of the central component is 0.81 mJy/beam. The high-resolution image shows the
beautiful ring oriented at a position angle of $\sim$135$^\circ$, with individual peaks of radio
emission. The extent of the radio ring is $\sim$9.1$\times$6.2 arcsec$^2$, which corresponds to 
a linear size of 1060$\times$720 pc$^2$. The positions and flux densities 
of the different components were estimated by fitting Gaussians and an underlying
tipped plane at the  peaks of radio emission 
which are at least $\sim$5 times higher than the rms noise level. These values are
subject to significant uncertainties because of difficulties in defining the
background flux density levels in the vicinity of each of the knots. Due to the
irregular background and limited angular resolution of the observations we could
not get satisfactory fits for four of the knots. All the components are listed in 
Table 1, along with their flux density estimates in units of mJy/beam. The 
components have been named such that the first four digits represent the seconds of
right ascension and the digits after the $+$ sign the arsec of declination.  We have 
listed the peak flux densities of the components from the image (column 2), as well as 
those obtained from the Gaussian fits (column 3). This gives an idea of the
contribution of the background at the different locations. The radio luminosities 
of the individual components are listed  in units of 10$^{19}$ W Hz$^{-1}$ using the
flux densities from the Gaussian fits, if available. The luminosities of those without 
satisfactory Gaussian fits have been estimated from the peak values in the image
and are enclosed in parentheses. The nuclear component has been marked with the 
superscript $n$.
 
The size and orientation of the ring seen at radio frequencies is roughly similar to that
seen at optical, H$\alpha$, infrared and mm wavelengths. The correspondence between the
positions of the radio peaks and those seen at other wavelengths is not precise 
although
they occur in the same general region. For example, by aligning the nuclear components,
it can be seen that in the north-western region, the radio knot 36.44+50.6 appears 
spatially coincident with a bright peak in the $B^{\prime}-K^{\prime}$ image (P\'{e}rez et 
al. 2000) as well as a prominent peak in the H$\alpha$ contours (Gonz\'{a}lez Delgado et al. 
1997). On the other hand, there are no such corresponding bright features for the nearby 
brighter radio components 36.19+50.0 and 36.02+49.1.  

The separation between the successive peaks of radio emission listed in Table 1 ranges
from $\sim$0.$^{\prime\prime}$8 to 3.$^{\prime\prime}$5. These separations are reasonably
uniformly distributed between 0.$^{\prime\prime}$8 and 2.$^{\prime\prime}$6, with only
the largest separation being beyond this range. There does not appear to  be a regular
period in the separation of the peaks of emission. If the radio peaks trace the star
forming regions, one would expect these to be regularly spaced in the gravitational
instability models (Elmegreen 1994).

The two most prominent regions of
radio emission are towards the north-west and the south-east. From CO (1-0) and
HCN (1-0) observations, Kohno et al. (1999) find that although both CO and HCN emission
are dominated by `twin-peaks' morphology, the HCN emission is better correlated
spatially with the peaks of radio emission, which are identified with the regions of
massive star formation. In this case, the peaks of  CO emission are identified with
the region of crowding of the $x_{\rm 1}$ and $x_{\rm 2}$ orbits.

\subsection{Nature of radio emission}

The radio emission from the circumnuclear region is possibly a mixture of thermal and 
non-thermal emission with the discrete components being either supernova
remnants (SNRs) or \hbox{H\,{\sc ii}} regions. The diagnostics which are usually employed 
to distinguish them are the brightness temperatures, as in the starburst galaxy Arp 220 
(Smith et al. 1998), and radio spectral indices. The radio components are usually too
weak for polarization measurements.

The brightness temperature of the components in NGC6951 is small, $\lapp$ 15$^\circ$ K, and
does not provide an unambiguous resolution of the nature of the compact components. 
Spectral index information on the individual knots is not available; however the spectral
index, $\alpha$ defined as S$\propto\nu^{-\alpha}$, of the entire circumnuclear emission 
estimated from scaled-array VLA observations between $\lambda$20 and 6 cm is 0.84 (HU). 
Convolving the $\lambda$3.6 cm image to lower resolutions and comparing with the images of
Saikia et al. (1994) and HU also suggests a non-thermal spectrum. 

However, a few caveats
need to be borne in mind while examining the nature of the emission of the components 
based on their spectra.
With inadequate resolution \hbox{H\,{\sc ii}} regions and SNRs may not be resolved from each other
(cf. Gordon et al. 1999).  Also, although the overall 
spectrum from coarser resolution might appear non-thermal, individual compact components 
identified from higher-resolution observations may be due to \hbox{H\,{\sc ii}} regions. An example of this
is the blue compact galaxy Henize 2-10 which is an archetypal starburst galaxy containing a
large population of Wolf-Rayet stars (Conti 1991). The integrated radio spectrum estimated
from observations largely with an angular resolution of about a kpc has a non-thermal spectrum,
with $\alpha\sim0.54$, while the compact radio knots observed with a resolution of $\sim$30 pc
have a flat or inverted spectrum (Kobulnicky \& Johnson 1999). These are identified with dense, 
compact \hbox{H\,{\sc ii}} regions which are strong sources of free-free emission.
For example, the features in Henize 2-10 (d$\sim$9 Mpc) have a luminosity of
$\sim$9$\times10^{18}$  W Hz$^{-1}$ at $\lambda$3.6 cm, electron densities of up to 5000 cm$^{-3}$
and harbour the youngest `super-star clusters' (Kobulnicky \& Johnson 1999). A similar feature
has been reported in the nuclear region of NGC5253 
(Turner et al. 1998, 2000)
which has a luminosity of $\sim$2$\times10^{19}$  W Hz$^{-1}$ at $\lambda$2 cm. Infrared
observations confirm it to be an \hbox{H\,{\sc ii}} region excited by a dense cluster of young
stars (Gorjian et al. 2001).

The nature of the compact radio components in other galaxies with either a 
circumnuclear ring or a nuclear starburst  might provide
some clues towards understanding the nature of the components in NGC6951. A number of these
galaxies have been observed at more than one frequency. Amongst the 
galaxies with radio rings, the radio spectrum of the ring in NGC613 was found to be 
$\sim0.65\pm0.10$ (Hummel \& J\"{o}rs\"{a}ter 1992), the components in NGC1097 (Hummel et al. 1987)
and NGC1365 (Saikia et al. 1994; Sandqvist et al. 1995; Stevens et al. 1999)
tend to have non-thermal spectra, as well as the ring of emission in the Seyfert galaxy
NGC7469 (Wilson et al. 1991). In NGC4736, which has a continuum ring, Duric \& Dittmar (1988) 
found 9 of their compact components to have a thermal spectrum between $\lambda$20 and 6 cm,
and the remaining 10 to have a non-thermal spectrum. Turner \& Ho (1994) observed this galaxy
at  $\lambda$6 and 2 cm, and found 9 of their 14 sources to have $\alpha<0.2$. 

The compact components in these systems are a mixture of both SNRs and \hbox{H\,{\sc ii}} regions occuring
in an area of high star formation.  The proportion of these two types of objects has 
been investigated in some detail in the archetypal starburst galaxies M82 
(Unger et al. 1984; Kronberg et al. 1985) and NGC253 (Antonucci \& Ulvestad 1988). 
Spectral studies of selected components in M82 (Wills et al. 1997; Allen \& Kronberg 1998) 
showed that of the 26 components, only one was identified with a thermal emission spectrum. However,
more recent observations of a further 20 largely weaker components suggest that only 5 of these
are SNRs while the remaining 15 are likely to be \hbox{H\,{\sc ii}} regions (McDonald 2001). Thus about 
two-thirds of the compact objects are identified with SNRs. In NGC 253
about a half of the sources with spectral information have flat spectra and are dominated
by emission from \hbox{H\,{\sc ii}} regions, while the remaining half
have steep spectra and are identified with SNRs (Ulvestad \& Antonucci 1991, 1997). A similar
situation is also seen in the nearby merger NGC4038/4039 (the Antennae), where about a third
of the objects have been suggested to be thermal and the remaining two-thirds non-thermal
(Neff \& Ulvestad 2000). The spectral indices of the components in NGC1808 tend to be nonthermal
(Saikia et al. 1990; Collison et al. 1994). In the starburst galaxy NGC2146 which has an
S-shaped structure with several knots of emission (Kronberg \& Biermann 1981; Saikia et al. 1994),
spectral index measurements suggest three sources to be good candidates for SNRs or radio 
supernovae, and 6 to be associated with compact, dense \hbox{H\,{\sc ii}} regions (Tarchi et al. 2000). 
Based on these studies of a wide variety of starburst galaxies, 
it is reasonable to assume that the ring in NGC6951 consists of a mixture of 
both these types of objects, although our linear resolution of about 100$\times$83 pc$^2$ 
is inadequate to identify individual SNRs and  \hbox{H\,{\sc ii}} regions. 

\subsection{Radio luminosity and supernova rate}

The compact components seen in NGC6951 have a median luminosity of 10$^{19}$ W Hz$^{-1}$
at  $\lambda$3.6 cm. The value is similar when one considers those with Gaussian fits as
well as for the entire list.  The luminosity distribution is skewed towards lower values,
with a tail extending towards higher luminosities, suggesting that we have not been able
to identify the weaker components from the existing observations.
 
It is interesting to compare the values obtained for NGC6951 with those obtained for other
galaxies exhibiting high star formation rates. For ease of comparison we have considered those
for which observations have been made at $\lambda$3.6 cm. The median luminosity of the
components in NGC253 (d$\sim$2.5 Mpc) observed with a linear resolution of $\sim$3 pc is
$\sim$4.5$\times10^{17}$  W Hz$^{-1}$ (Ulvestad \& Antonucci 1991), while those in M82 (d$\sim$3.4
Mpc) observed with a similar linear resolution has a median value of 
$\sim$2$\times$10$^{18}$ W Hz$^{-1}$ (Huang et al. 1994; Allen \& Kronberg 1998). 
Considering more distant galaxies, the median luminosity for the archetypal starburst 
galaxy in the southern hemisphere, NGC1808 (d$\sim$10.9 Mpc), is about
8$\times10^{18}$  W Hz$^{-1}$ with a linear resolution of 30 pc (Collison et al. 1994), 
while the knots in NGC1365 (d$\sim$20 Mpc) has a median luminosity of 
$\sim$3.5$\times10^{19}$  W Hz$^{-1}$ with a linear resolution of about a 100 pc
(Stevens et al. 1999). The increase in luminosity for the more distant systems
is largely due to the coarser linear resolution of the observations. To examine this we
have convolved a multi-configuration VLA image of M82 with a linear resolution of $\sim$6 pc
to a resolution comparable with our observations of NGC6951. In the convolved image, the individual
components are indistinguishable and the peak brightness has increased by a factor of $\sim$16,
corresponding to a luminosity of 6.1$\times10^{20}$ W Hz$^{-1}$ at 5 GHz. This is more luminous
than the individual radio components in NGC6951. A more detailed comparison would require
observations of NGC6951 with higher linear resolution.

We estimate the SN rate in NGC6951 using the model of Condon \& Yin (1990) which was based
on data for our Galaxy. Assuming that the entire emission at 1465 MHz (Hummel et al. 
1985) is non-thermal, we estimate the SN rate to be $\sim$0.07 yr$^{-1}$. 
Bearing in mind that the emission
at 1465 MHz could also contain a thermal contribution, the SN rate in NGC6951 appears 
to be reasonably consistent with 
estimates of $\sim$0.1 yr$^{-1}$ for M82 (e.g. Huang et al. 1994), $\lapp$0.1 to 0.3 for 
NGC253 (Ulvestad \& Antonucci 1997), $\sim$0.1 yr$^{-1}$ for the clumpy irregular 
starburst galaxy Mrk 325 (Condon \& Yin 1990) and also for the starburst galaxy NGC3448 of
the Arp 205 system (Noreau \& Kronberg 1987).

\subsection{The central component}

By fitting a two-dimensional gaussian  and an underlying tipped plane 
to the central component, we find it to be
resolved with a deconvolved angular size of 0.7$\times$0.2 arcsec$^2$ along a PA of 
156$^\circ$. This corresponds to a linear dimension of $\sim$80$\times$20 pc$^2$. HU
also find evidence of a similar extension but are unable to confirm it because of the
coarser resolution of their observations with the restoring beam being elongated along
a similar PA.  The H$\alpha$ image of Knapen (1996) also seems to show a similar extension. 
The spectral index of the nuclear component obtained by smoothing the
$\lambda$3.6 cm image to that of the lower-resolution $\lambda$6 cm image of Saikia et al. (1994) is
$\sim$0.6. The scaled-array observations of HU yield slightly higher values of 0.69 and 0.90
between $\lambda$20 and 6 cm for the peak and integrated flux densities respectively. 
The supernova rate estimated from the nuclear
radio flux density (cf. Colina \& P\'{e}rez-Olea 1992) is about 0.003 yr$^{-1}$, which
is much smaller than the SN rate in bright nuclear starbursts. As pointed out
by P\'{e}rez et al. (2000), the low SN rate is consistent with the faintness of the
nucleus in H$\alpha$. The resolved central component, which is elongated and has a 
steep spectral index, could be due to a small-scale, weak nuclear jet from the Seyfert/LINER nucleus. 

\section{Conclusions}
High-resolution radio observations of the circumnuclear region of NGC6951 reveal a ring of radio
emission with many discrete components and a slightly resolved nuclear component. The discrete
components have a median luminosity of 10$^{19}$ W Hz$^{-1}$ when observed with a linear resolution 
of $\sim$90 pc, and are likely to consist of a mixture of both SNRs and \hbox{H\,{\sc ii}} regions. 
The SN rate has been estimated to be $\sim$0.07 yr$^{-1}$, which is 
reasonably consistent with those found in some
of the archetypal starburst galaxies. The central component, which is resolved and has a 
steep radio spectrum, could be due to a small-scale jet from the active nucleus.

\begin{acknowledgements}
The National Radio Astronomy Observatory is a facility of the National Science Foundation
operated under co-operative agreement by Associated Universities, Inc.  We have made use of the
NASA/IPAC Extragalactic Database (NED), which is operated by the Jet Propulsion Laboratory,
California Institute of Technology under contract with the National Aeronautics and Space
Administration. We thank Jim Ulvestad for several comments and suggestions which have improved the paper 
significantly. We also thank Graham Smith and Peter Thomasson for their detailed comments on
the manuscript.  One of us (DJS) would like to thank the Cave Lecture Fund and the 
Principal's Development Fund of Queen's University, Kingston, and Director, Jodrell Bank Observatory
for financial support, and Judith Irwin, Dieter Brueckner and Peter Thomasson for hospitality during 
different phases of this piece of work. 
\end{acknowledgements}

\end{document}

%% file: table1.tex
\begin{table}

\caption{The compact components}
\begin{tabular}{l c c c}
\hline
Source   & S$_{image}$ & S$_{gauss}$ & Lum       \\ 
         &                &                & $\times$10$^{19}$ \\
         & mJy/b          & mJy/b          & W/Hz \\
\hline
36.02+49.1   &  0.234 & 0.23 & 1.60 \\   
36.05+46.3   &  0.116 & 0.10 & 0.69 \\ 
36.07+47.8   &  0.205 & 0.20 & 1.39 \\
36.16+45.2   &  0.125 & 0.07 & 0.49 \\
36.19+50.0   &  0.270 & 0.24 & 1.67 \\
36.44+43.6   &  0.097 & 0.08 & 0.56 \\ 
36.44+50.6   &  0.157 & 0.14 & 0.97 \\
36.64+46.3$^n$  &  0.536 & 0.54 & 3.75 \\
36.82+49.2   &  0.105 & 0.09 & 0.63  \\
36.96+42.5   &  0.109 &      & (0.76)\\
36.98+48.4   &  0.128 & 0.11 & 0.76  \\
37.08+42.3   &  0.145 &      & (1.01)\\
37.12+45.8   &  0.166 &      & (1.15)\\
37.16+43.0   &  0.202 &      & (1.40)\\
37.20+44.7   &  0.335 & 0.30 & 2.08  \\

\hline
\end{tabular}
\end{table}